\documentclass[runningheads]{llncs}
\usepackage[american]{babel}
\usepackage[utf8]{inputenc}
\usepackage[T1]{fontenc}

\usepackage{cite}
\usepackage{amsmath,amssymb,amsfonts}
\usepackage{algorithm}
\usepackage{algpseudocode}
\usepackage{graphicx}
\usepackage{textcomp}
\usepackage{booktabs}
\usepackage{lipsum}
\usepackage[normalem]{ulem}

\usepackage{url}
\usepackage{hyperref}
 \hypersetup{
 	bookmarksdepth=-2
 } 

\usepackage{multirow,subfig}
\usepackage{makecell}  

\begin{document}

\title{SPH-EXA: Enhancing the Scalability of SPH codes Via an Exascale-Ready SPH Mini-App}

\authorrunning{Guerrera D. and Cavelan A.}
\titlerunning{SPH-EXA Mini-App}

\author{Danilo Guerrera\inst{1} \and Aurélien Cavelan\inst{1} \and Rubén M. Cabezón\inst{2}
\and David Imbert\inst{3} and Jean-Guillaume Piccinali\inst{4}  \and Ali Mohammed\inst{1}
\and Lucio Mayer\inst{5} \and Darren Reed\inst{5} \and Florina M. Ciorba\inst{1}}

\institute{Department of Mathematics and Computer Science, University of Basel, Switzerland\\
\email{\{firstname.lastname\}@unibas.ch},
\and
Scientific Computing Center (sciCORE), University of Basel, Switzerland,\\
\email{ruben.cabezon@unibas.ch},
\and
XXX Software, YYY, ZZZ,\\
\email{contact@xxx.com},
\and
Scientific Computing Support, Swiss National Supercomputing Centre, Lugano, Switzerland,\\
\email{jgp@cscs.ch},
\and
Center for Theoretical Astrophysics and Cosmology, Institute for Computational Science, University of Z{\"u}rich, Switzerland,\\
\email{\{lmayer,reed\}@physik.uzh.ch}
}

\maketitle

\begin{abstract}
Numerical simulations of fluids in astrophysics and computational fluid
dynamics (CFD) are among the most computationally-demanding calculations, in
terms of sustained floating-point operations per second, or FLOP/s. It is
expected that these numerical simulations will significantly benefit from the
future Exascale computing infrastructures, that will perform $10^{18}$ FLOP/s.
The performance of the SPH codes is, in general, adversely impacted by several
factors, such as multiple time-stepping, long-range interactions, and/or
boundary conditions. In this work an extensive study of three SPH
implementations SPHYNX~\cite{SPHYNX}, ChaNGa~\cite{CHANGA}, and
XXX~\cite{XXX} is performed, to gain insights and to expose any
limitations and characteristics of the codes. These codes are the starting
point of an interdisciplinary co-design project, SPH-EXA, for the development
of an Exascale-ready SPH mini-app. We implemented a rotating square
patch~\cite{SqPatch} as a joint test simulation for the three SPH codes and
analyzed their performance on a modern HPC system, Piz Daint.
The performance profiling and scalability analysis conducted on the three
parent codes allowed to expose their performance issues~\cite{WRAP18}, such as
load imbalance, both in MPI and OpenMP. Two-level load balancing has been
successfully applied to SPHYNX to overcome its load imbalance. The performance
analysis shapes and drives the design of the SPH-EXA mini-app towards the use
of efficient parallelization methods, fault-tolerance mechanisms, and load
balancing approaches.  \end{abstract}

\section{Introduction}

\textit{Faster, Higher, Stronger} is the Olympic motto proposed by Pierre de
Coubertin upon the creation of the Olympic Committee in 1894. One hundred and
twenty-five years later, this motto's fundamental idea represents somehow the
computational effort done in high-performance computing and its scientific
applications in the last decades, although re-phrased to \textit{Bigger,
Faster, Longer}. Numerical simulations grow more complex and, as a consequence,
more computationally demanding, while addressing more detailed and intricated
scientific questions. One paradigmatic example of this can be found in
Computational Astrophysics. This is a field where our ability to understand
complex processes, like those that form the stars or make them explode as
Supernovas, is not only limited by our incomplete knowledge of the underlying
physics, but also by our capability to harness the current computational power.
\textit{Bigger} simulations are needed to include all relevant processes for
the scenario at study. This implies that they should be \textit{faster} to
compensate for the additional computational expense that detailed physics
entail. Finally, we are usually interested in \textit{longer} simulations, that
allow a consistent study of a process and its consequences ab initio, but also
in terms of \textit{resilience}, so that the simulations can perform their
calculations accurately even when external factors, like silent data
corruption, occur.

The SPH-EXA project was born with these ideas in mind: developing a mini-app to
perform Smoothed Particle Hydrodynamics (SPH) simulations that are
\textit{Bigger, Faster, Longer}. To that extent, we brought together a group of
Computational Scientists, Astrophysicists, and Fluid Dynamics experts to share
their respective knowledge and find a common ground where the SPH-EXA mini-app
could be built, targeting the upcoming Exascale infrastructures. In this
respect, this is a co-design project in terms of interdisciplinarity.

\begin{itemize}
\item Contribution details
\end{itemize}
Interdisciplinary co-design and co-development is the recommended approach for developing a mini-app, to leverage the involvement of the developers of the parent codes in the mini-app design and implementation process~\cite{HDCW09}. The SPH-EXA project follows such recommendation, as both computer scientists and computational scientists (that developed the three parents codes) are part of the team working on the \mbox{SPH-EXA} mini-app. The development has thus followed a building blocks approach, with the resulting mini-app being shaped while adding new features both on the SPH side (e.g., SPH operands) and the HPC side (e.g., adopting a specific data structure or parallel algorithm).
Lighter than production codes, mini-apps are algorithm oriented and allow
modifications and experiments \cite{BVH12}. Replacing a data structure or a
specific algorithm can be easily done without additional cost. Immediate
results and computation time obtained by the mini-app allow anyone to make is
own idea about the efficiency of one or an other algorithm.

Mini-apps have been a privileged area for high performance computing
\cite{BVH12,HDCW09}. 
single function can be evaluated using different strategies leading to
different performances, even if the physical result is still the same. These
evaluation strategies may rely on vectorisation, node level multi-threading,
cross-node parallelism. Their efficiency also depends on platforms
configuration: presence of accelerators, CPU generation, interconnection
network. Therefore, a mini-app is a perfect portable sandbox to optimize a
numerical method such as the SPH method. 

Mini-apps also help testing new technical features which could not be available
in large production codes. For example, the three SPH codes are implemented
with different languages, are built with different compilers versions, rely on
different dependencies, and targets different operating systems and platforms.
Therefore, last SIMD instructions, last C++ standard, or a new library, may not
be tested in all production codes without a high development cost to adapt
these codes. A freshly designed mini-app does not suffer from maintenance
costs and allows estimating potential benefits of new features before
implementing them in production context.

Parts of the mini-app can be organized as independent libraries to help
integration of new features in production code. This organization avoids
implementing two times an interesting feature: one time in the mini-app, one
other time in the production code. Interfaces to this library should be
generic enough to work with any production code.

The mini-app introduced in this work is still in its early stages and offers limited capabilities for SPH simulations. As it has been designed with the goal to allow interfacing with production code and as a testing ground for new features, it is expected to benefit from the ideas and interest of the SPH community. A request for a new test cases could turn into the development of a missing SPH-related functionality and allow the exploration of alternative solutions on the HPC side.

\section{Related Work}
\label{sec:related-work}
Mini-apps or proxy-apps have received great attention in recent years, 
with several projects being developed or under development. 
The Mantevo Suite~\cite{HDCW09}, from the high performance computing~(HPC) community, 
and developed at Sandia National Laboratory is one of the first large \mbox{mini-app} set. 
It includes \mbox{mini-apps} that represent the performance of \mbox{finite-element} codes, 
molecular dynamics, and contact detection, to name a few. 

Another example is CGPOP~\cite{CGPOP}, a \mbox{mini-app} from the oceanographic community, 
that implements a conjugate gradient solver to represent the bottleneck of the full Parallel 
Ocean Program application.
CGPOP is used for experimenting new programming models and to ensure performance portability 

At Los Alamos National Laboratory, MCMini~\cite{MCMini} was developed as a \mbox{co-design} 
application for Exascale research.
MCMini implements Monte Carlo neutron transport in OpenCL and targets accelerators and 
coprocessor technologies.

The CESAR Proxy-apps~\cite{CESAR} represent a collection of mini-apps belonging to three main 
fields: thermal hydraulics (for fluid codes), neutronics (for neutronics codes), and coupling 
and data analytics (for \mbox{data-intensive} tasks).

One of the motivation behind the European ESCAPE project~\cite{ESCAPE} is to
define and encapsulate the fundamental building blocks (`Weather \& Climate
Dwarfs') that underlie weather and climate services. This serves as a
prerequisite for any subsequent co-design, optimization, and adaptation
efforts. One of the ESCAPE outcomes is Atlas~\cite{Atlas}, a library for
numerical weather prediction and climate modeling, with the primary goals to
exploit the emerging hardware architectures forecasted to be available in the
next few decades. Interoperability across the variegated solutions that the
hardware landscape offers is a key factor for an efficient software and
hardware co-design~\cite{Schulthess2015}, thus of great importance when
targeting Exascale systems.

Similar to these works, the creation of a mini-app directly from existing codes
rather than the development of a code that mimics a class of algorithms has
been recently discussed~\cite{Messer2015}. A scheme to follow was proposed
therein that must be adapted according to the specific field the parent code
originates in. To maximize the impact of a mini-app on the scientific
community, it is important to keep the build and run system easy enough, to not
discourage potential users. The building should be kept as simple as a
Makefile and the preparation of the run to a handful of command line arguments:
``\textit{if more than this level of complexity seems to be required, it is
possible that the resulting MiniApp itself is too complex to be
human-parseable, reducing its usefulness.}" \cite{Messer2015}. The present
work introduces the interdisciplinary co-design of an SPH-EXA mini-app with
three parent SPH codes originating in the astrophysics academic community and
the industrial CFD community.
This represents a category not discussed in~\cite{Messer2015}. 

Skeleton applications, the name used to refer to reduced versions of
applications that produce the same network traffic of the full ones, are of
interest to model the performance of networks through simulation. An
auto-skeletonization approach has recently been proposed to auto-skeletonize a
full application via compiler pragmas: an interesting idea to produce flexible
skeletons, that serve as a tool to study balanced Exascale interconnect
designs~\cite{Wilke2018}. Such a skeletonization approach may be of interest in
annotating a full application to obtain a mini-app that reflects exactly the
corresponding production code.

\section{Co-Design of the Mini-App}

Being optimization critical to achieve the scalability needed to exploit
Exascale computers, the long-term goal of SPH-EXA~\cite{SPHEXA} is to provide a
parallel, optimized, state-of-the-art implementation of basic SPH operands with
classical test cases used by the SPH user community. This can be implemented at
different levels: employing state-of-the art dynamic load balancing algorithms,
fault-tolerance techniques, programming languages, tools, and libraries.

In reaching this goal, interdisciplinary co-design and co-development is the
recommended approach for developing a mini-app, to leverage the involvement of
the developers of the parent codes in the mini-app design and implementation
process~\cite{HDCW09}. 

Thanks to the co-design and co-development, it is straightforward in the
analysis phase to individuate the common and best features of the
state-of-the-art parent codes, with a focus on the physics therein, highlighted
in Table~\ref{table:differences_scientific}. In the first stage the focus has
been on implementing a vanilla SPH solver, representing the basis of the future
mini-app. The momentum equation implemented can be found in \cite{rosswog2009}, 
while the artificial viscosity has been derived from \cite{monaghan1983}.
To calculate the value of the next timestep the Courant condition is applied; the 
updates of the positions and the energy follow a Press and 
Adams-Bashforth method, respectively.
While not all existing techniques and algorithms need to be
implemented, some of them, such as the SPH interpolation kernels, can be later
developed as separate interchangeable modules.
In that sense, Table~\ref{table:differences_scientific} shows also the domain
science techniques and algorithms that the mini-app features.

Simultaneously to the effort on the SPH solver implementation, HPC related
aspects regarding the development of the code have been addressed. The solver
has evolved from being sequential, to having intra-node parallel capabilities
via OpenMP, to inter-node parallelism via MPI.

The goal is to provide a reference implementation in MPI+X. MPI is the de facto
standard for HPC applications, due to the lack of a valid alternative for the
inter-node communication. The MPI+OpenMP programming model represents a
starting point, since it does not fully exploit the heterogeneous parallelism
in the newest architectures. While OpenMP 4.5~\cite{OpenMP} offers support for
accelerators, in the meanwhile other languages directly targeting accelerators
have being proposed and accepted by the community, such as OpenACC (a
directive-based programming model targeting a CPU+accelerator system, similar
to OpenMP), CUDA (an explicit programming model for GPU accelerators) and
OpenCL. In programming models, research has been focusing on the efficient use
of intra-node parallelism, able to properly exploit the underlying
communication system through a fine grain task-based approach, ranging from
libraries (Intel TBB~\cite{TBB}) to language extensions (Intel Cilk
Plus~\cite{CilkPlus} or OpenMP), to experimental programming languages with
focus on productivity (Chapel~\cite{Chapel}). Kokkos~\cite{Kokkos} offers a
programming model, in C++, to write portable applications for complex manycore
architectures, aiming for performance portability. HPX~\cite{HPX} is a
task-based asynchronous programming model that offers a solution for
homogeneous execution of remote and local operations. It is planned to port the
mini-app to at least one of the paradigms described, between the ones described
previously to explore their efficiency and potential on Exascale ready
machines.

Then, Table~\ref{table:differences_cs} reveals the computer science-related
similarities and differences of the codes, i.e., algorithms, techniques, and
other implementation-specific choices. Each code has a different history, has
been used for different purposes, and, therefore, uses different approaches in
its simulations. For example, all applications use standard checkpoint/restart
mechanisms to enable fault-tolerance when executing at scale. Yet, they all
use different domain-decomposition methods and scheduling techniques. It is
important to note that such features can dramatically affect the scalability of
the application, as shown in Section~\ref{sec:experiments}.

\begin{table*}[t]
\centering
\caption{Differences and similarities between SPHYNX, ChaNGa, XXX, and the SPH-EXA mini-app} 
\label{table:differences_scientific}
\resizebox{\textwidth}{!}{%
\begin{tabular}{@{}lllllllll@{}}
\toprule
 \textbf{SPH} & \textbf{Code}       & \multirow{2}{*}{\textbf{Kernel}}  & \textbf{Gradients}     & \textbf{Volume} & \textbf{Mass of}  & \textbf{Time-} & \textbf{Neighbour} & \multirow{2}{*}{\textbf{Self-Gravity}} \\
\textbf{Code } & \textbf{Version}       &  & \textbf{Calculation}     & \textbf{Elements} & \textbf{Particles}  & \textbf{Stepping} & \textbf{Discovery} &                                  \\ \midrule
SPHYNX   & 1.3.1 & Sinc     & IAD                       & Generalized     & Equal or Variable             & Global        & Tree Walk            & Multipoles (4-pole) \\
ChaNGa   & 3.3 & \makecell[l]{Wendland,\\ M4 spline} & Kernel derivatives & Standard               & Equal or Variable           & Individual    & Tree Walk            & Multipoles (16-pole)\\
XXX & 17.6 & Wendland & Kernel derivatives  & Standard   & Equal or Adaptive & Global    & Tree Walk      & No         \\
mini-app & 0.5 & \makecell[l]{Sinc, Wendland}  & \makecell[l]{Kernel derivatives, \\ IAD (in developement)}   & \makecell[l]{Standard, \\ Generalized (in developement)}       & \makecell[l]{Equal or Variable}               & Global       & Tree Walk            & \makecell[l]{Multipoles (16-pole)\\ (in developement)}\\
\bottomrule
\end{tabular}
} 
\end{table*}

\begin{table*}[t!]
\centering
\caption{Different and similar computer science-related aspects between SPHYNX, ChaNGa, XXX, and the SPH-EXA mini-app}
\label{table:differences_cs}
\resizebox{\textwidth}{!}{
\begin{tabular}{@{}lllcrllr@{}}
\toprule
 \textbf{SPH}        & \textbf{Domain}  		& \textbf{Load}     		& \textbf{Checkpoint-} & \multirow{2}{*}{\textbf{Precision}}  & \multirow{2}{*}{\textbf{Language}} & \multirow{2}{*}{\textbf{Parallelization}} & \multirow{2}{*}{\textbf{\#LOC}}  \\
\textbf{Code }        & \textbf{Decomposition} & \textbf{Balancing}     & \textbf{Restart}				 & 				  & 				& &            \\ \midrule
SPHYNX   & Straightforward     		& None (static)                      	   & Yes 		    	& 64-bit     & Fortran 90,  & MPI+OpenMP      & 25,000             \\
ChaNGa   & Space Filling Curve 		& Dynamic			&  Yes          		 & 64-bit    & C++  & MPI+OpenMP+CUDA & 110,000 \\
XXX & Orthogonal Recursive Bisection & Local-Inner-Outer  		& Yes               	& 64-bit  & \makecell[l]{Fortran 90} & MPI  & 37,000 \\
mini-app & HTree based  & None (static)  & No  & 64-bit  & C++ &  MPI+OpenMP & 3,500  \\
\bottomrule
\end{tabular}
} 
\end{table*}


In its current status, the mini-app provides two different implementations for the SPH kernel, namely Sinc, whose benefits are described in~\cite{SINC}  and Wendland, originally presented in~\cite{Wendland}. The Sinc functions has been defined as:
\begin{equation}
W^s_n(v,h,n)=\frac{B_n}{h^d}S_n(\frac{\pi}{2}v) \text{for $0 \ge v \le 2$,}
\end{equation}
\label{eq:sinc}
where $S_n(.) = sinc_n(.)$, and where n is the index of the kernel and $B_n$ a normalization constant. The function $sinc(\frac{\pi}{2}v)= \frac{sin(\frac{\pi}{2}v)}{\frac{\pi}{2}v}$, widely used in spectral theory, has been chosen.

The Wendland kernel instead, has been defined as:
\begin{equation}
W(v,h,n)=
\begin{cases}
\sigma_W(1-\frac{v}{2})^4(2v+1) &\text{for $0 \ge v \le 2$,}\\
0 &\text{for $v > 2$,}
\end{cases}
\end{equation}
\label{eq:sinc}
where the normalization $\sigma_W$ is $21/(16\pi)$.

Additional kernels can be plugged in, upon implementation (in the above cases 6 LOC only), proving how such a mini-app can result useful in allowing easy and quick modification to experiment alternative solutions.

On current HPC systems it is common to have more than one programming environment installed .
On Piz Daint (at least) 4 compiler suites (Intel,
GNU, PGI and Cray) are available to the users together with communication and
scientific libraries tuned by Cray and CSCS. As the programming environment and
operating system gets updated regularly, it is considered good practice to
continuously test that a code repeatedly builds and runs correctly independently
of any given change to the software stack and/or code base. In order to achieve
this goal of continuous integration, all involved steps of the project have
been designed to run fully automated with \href{https://user.cscs.ch/tools/continuous/}{Jenkins}.
As a result, the mini-app has been tested with the following compilers: gcc/6.2.0, 
gcc/7.3.0, gcc/8.2.0, intel/18.0, intel/19.0, cce/8.6.1, cce/8.7.6, pgi/18.7 and 
clang/7.0.0. While continuous testing gives information about the current state of
the code, it also allows to quickly pinpoint sources of problems and to apply
changes when needed. History on the information about previous failures or
success can easily be accessed on the Jenkins dashboard. The regression tests for
the mini-app have been generated using the
\href{https://github.com/eth-cscs/reframe}{ReFrame} framework.  


 Jenkins is a self-contained, open source automation server which can be used
 to automate all sorts of tasks related to building, testing, and delivering or
 deploying software.

%
 ReFrame is a framework for writing regression tests for HPC systems. The goal
 of this framework is to abstract away the complexity of the interactions with
 the system, separating the logic of a regression test from the low-level
 details, which pertain to the system configuration and setup. This allows users
 to write easily portable regression tests, focusing only on the functionality.
 
 Regression tests in ReFrame are simple Python classes that specify the basic
 parameters of the test. The framework will load the test and will send it down
 a well-defined pipeline that will take care of its execution. The stages of
 this pipeline take care of all the system interaction details, such as
 programming environment switching, compilation, job submission, job status
 query, sanity checking and performance assessment.
 
 Writing system regression tests in a high-level modern programming language,
 like Python, poses a great advantage in organizing and maintaining the tests.
 Users can create their own test hierarchies, create test factories for
 generating multiple tests at the same time and also customize them in a simple
 and expressive way.
Of the three SPH codes, two are in Fortran 90 (SPHYNX and XXX) and one in
C++ (ChaNGa); note that the future version of XXX in development is in C++.
As a compiled static language, both C++ and Fortran allows a close management
of memory and higher performance than Python, Matlab, or R \cite{Cad17}. 

Fortran is a procedural language suffering from severe maintenance issues:
global variables, complex divisions, interlinking. Therefore, Fortran code is
hard to modify, unit test, and extend \cite{Cad17}. Fortran 2003 goal was to
become object-oriented; but in 2017, the 2003 standard was still not completely
supported by all compilers \cite{For17}. On the contrary, C++ is close to the
hardware and also completely supports object-oriented \cite{Cad17}. That is
why C++ has been chosen as the mini-app language.

C++ also supports header-only libraries which does not require a software
factory. Header files just have to stand near the client "cpp" files.  This is
an asset to spread more easily the mini-app in the community. 

Moreover, header-only libraries can completely benefit from template meta
programming to perform operations or detect errors during compile-time instead
of runtime. This also leads to higher runtime performance \cite{Mey05}.
Template meta programming allows generating custom code based on policy
choices.
The mini-app introduced in this work is still in its early stages and offers
limited capabilities for SPH simulations. As it has been designed with the goal
of allowing an easy interface with production codes and as a testing ground for
new features, it is expected to benefit from the ideas and interests of the SPH
community. A request for a new test case could turn into the development of a
missing SPH-related functionality and allow the exploration of alternative
solutions on the HPC side.

\section{Status of the Mini-App}
\label{sec:experiments}

The complexity of the scenarios simulated in CFD and Astrophysics usually
forbids the possibility to perform simulations with continuously increased
resolution and different codes, so that a convergence to zero differences on
the results can be found. Often, it is neither possible nor reasonable to
obtain sufficient computational resources to perform simulations that are
``converged'' throughout the computational domain in a mathematical sense. It
is much more important to limit the deviations in under-resolved regimes by
enforcing fundamental conservation laws. Additionally, the emergence of
stochastic processes, like turbulence, renders the pursuit for mathematical
convergence impossible, yet still constrained by conservation laws. As a
consequence, overall physics properties of the simulated scenarios remain
robust, even if slightly different results are obtained when using different
codes to ``solve'' the same set of equations. Therefore, comparing results of
different hydrodynamical codes to the same initial conditions has been proved
to be highly beneficial to gain understanding in complex scenarios, in the
behavior of the codes, and to discover strengths and weaknesses of those.
These comparisons are not uncommon in CFD and
Astrophysics~\cite{liebendoerfer2005, agertz2007, tasker2008, cabezon2018}.
The SPH codes comparison presented herein is in this spirit, yet with a focus
on performance-related aspects from computer science.

\begin{table*}
\caption{Test simulations and their characteristics}
\label{table:differences_simulations}
\resizebox{\textwidth}{!}{
\begin{tabular}{@{}llcccr@{}}
\toprule
 \multirow{2}{*}{\textbf{Test Simulation}}        	  & \multirow{2}{*}{\textbf{Description}}	&  \multirow{2}{*}{\textbf{Domain Size}}    		& \textbf{Simulation} &  \multirow{2}{*}{\textbf{SPH Code}}  & \textbf{Test} \\
								        & 								 & 									& \textbf{Length}	& 							 & \textbf{Platform} \\ \midrule
\makecell[l]{Rotating Square Patch~\cite{SqPatch}}   & Rotation of a free-surface 	&  3D, $10^{6}$ particles  &  40 time-steps    	&  \makecell[l]{SPHYNX, ChaNGa \\ XXX, mini-app}& Piz Daint\\
							      \bottomrule
\end{tabular}
} 
\vspace{-0.5\baselineskip}
\end{table*}

The common test case chosen to validate the results obtained through the
mini-app against the ones of the parent codes is the Rotating Square Patch.

This test was first proposed by~\cite{SqPatch} as a demanding scenario
for SPH simulations. The presence of negative pressures stimulates the
emergence of unphysical tensile instabilities that destroy the system.
Nevertheless, these can be suppressed either using a tensile stability control
or increasing the order of the scheme~\cite{SPHFLOW}. As a consequence, this
is a commonly used test in CFD to verify hydrodynamical codes, and it is
employed in this work as a common test for the three codes.

The setup here is similar to that of \cite{SqPatch}, but in 3D. The
original test was devised in 2D, but the SPH codes used in this work normally
operate in 3D. To use a test that better represents the regular operability of
the target codes, the square patch was set to $[100 \times 100]$ particles in
2D and this layer was copied 100 times in the direction of the Z-axis. This
results in a cube of $10^6$ particles that, when applying periodic boundary
conditions in the Z direction, is equivalent to solving the original 2D test
100 times, while conserving the 3D formulation of the codes. The initial
conditions are the same for all layers, hence they depend only on the X and Y
coordinates. The initial velocity field is given such that the square rotates
rigidly:
\begin{equation}
v_x(x,y)=\omega y;\qquad v_y(x,y)=-\omega x,
\label{eq:square_v0}
\end{equation}
\noindent
where $v_x$ and $v_y$ are the X and Y coordinates of the velocity, and $\omega = 5$~rad/s is the angular velocity. 
The initial pressure profile consistent with that velocity distribution can be
calculated from an incompressible Poisson equation and expressed as a rapidly
converging series:
%
\begin{multline}{\notag}
P_0=\rho\sum_{m=0}^\infty\sum_{n=0}^\infty\frac{-32\omega^2}{mn\pi^2\left[\left(\frac{m\pi}{L}\right)^2 + \left(\frac{n\pi}{L}\right)^2 \right]} \times  \\
\sin\left(\frac{m\pi x}{L}\right)\sin\left(\frac{n\pi x}{L}\right),
\end{multline}
\label{eq:square_p0}
%
\noindent
where $\rho$ is the density and $L$ is the side length of the square.

\subsection{Experimental Results}

\subsubsection{System overview - Piz Daint}

The experiments were performed on the hybrid partition of the Piz
Daint\footnote{\url{https://www.cscs.ch/computers/piz-daint/}} supercomputer
using PrgEnv-Intel, Cray MPICH 7.7.2 and OpenMP 4.0 (version/201611).
At the time of writing, this
supercomputer consisted of a multi-core partition of more than 1,600 Cray XC40 nodes with
two Intel Xeon E5-2695 v4 (codename Broadwell) processors, which were not used
in this study, as well as the hybrid partition of more than 5,000 Cray XC50 nodes. These
hybrid nodes are equipped with an Intel E5-2690 v3 CPU (codename Haswell) and a
PCIe version of the NVIDIA Tesla P100 GPU (codename Pascal) with 16 GB second
generation high bandwidth memory (HBM2). The nodes of both partitions are
interconnected in one fabric based on Aries technology in a Dragonfly
topology\footnote{\url{http://www.cray.com/sites/default/files/resources/CrayXCNetwork.pdf}}.

\subsubsection{Runtime evaluation}
\begin{figure}[!htb]
	\centering
		\includegraphics[scale=0.7]{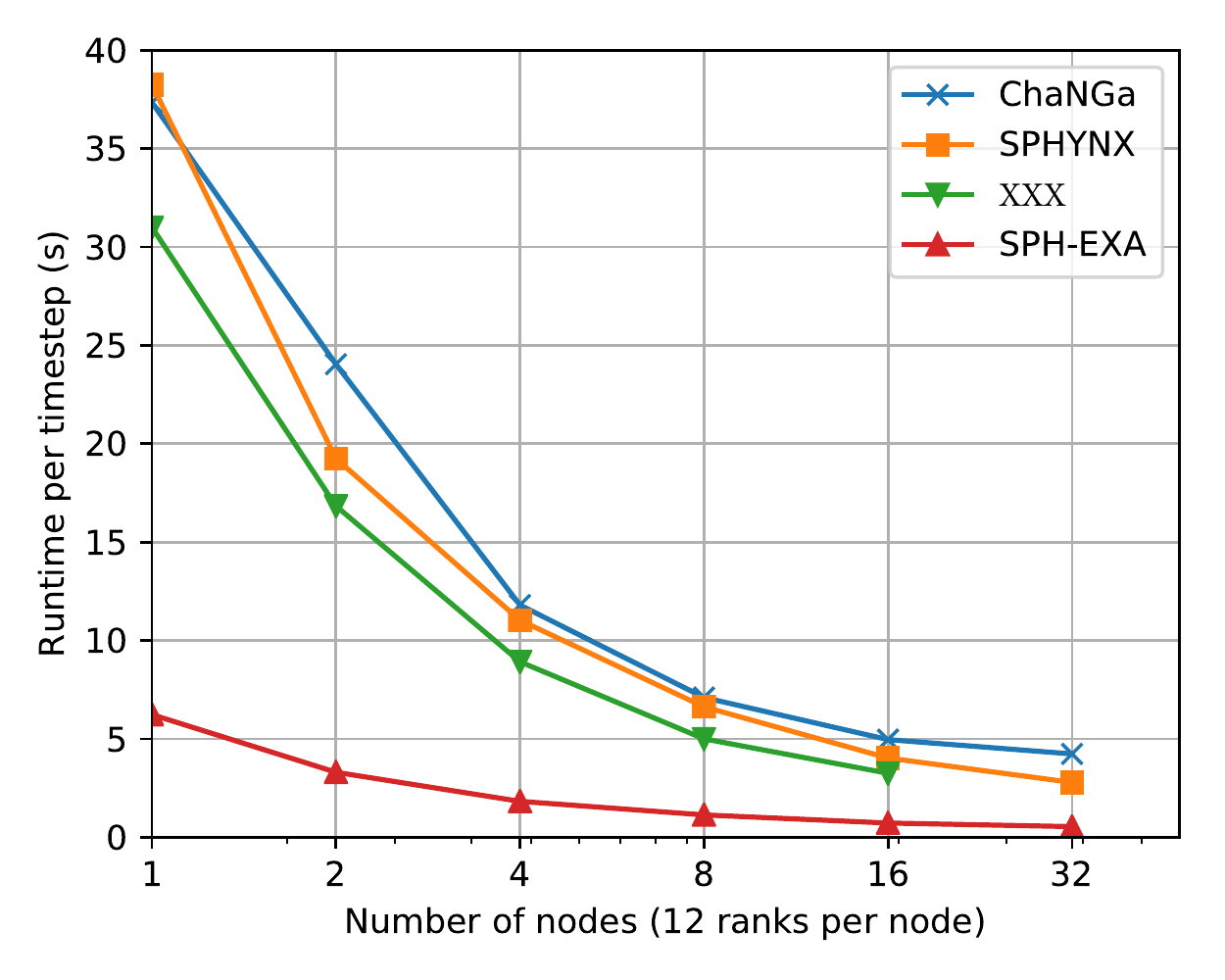}
	\caption{Average execution time per timestep (out of 40 timesteps) of ChaNGa (12 OpenMP tasks and 1 MPI rank per node), SPHYNX (12 OpenMP tasks and 1 MPI rank per node), XXX (12 MPI ranks per node) and SPH-EXA mini-app (12 MPI ranks per node), in function of the number of nodes used (1 node = 12 cores) on Piz Daint.}
	\label{fig.runtime}
\end{figure}

Figure~\ref{fig.runtime} shows the average execution time per timestep (in seconds), out of 40 time-steps, on Piz Daint up to 32 nodes (i.e., 384 cores).
The SPH-EXA mini-app is able to complete a timestep circa 6 times faster than the production codes, with different number of nodes.
Indeed, the proposed mini-app has been designed from scratch with performance in mind. The straightforward SPH implementation, together with the minimal and modern code style lead to increased performance compared to the legacy codes.
However, the SPH-EXA mini-app is currently limited to the bare SPH implementation.


\subsubsection{Strong scaling evaluation}
The measure of parallel scalability indicates how efficently an application is in utilizing an increasing number of processing elements. 
In the case of strong scaling the problem size is fixed, while the number of processing elements increases.
Its goal is to find a ``sweet spot", where the computation completes in a reasonable amount of time, while not wasting too many 
cycles due to the parallel overhead. The applications exhibit good strong scaling up to 16 compute nodes.
The scaling stalls when there are not enough particles/core (typically $10^4$) to keep the compute nodes busy.
The measured global efficiency decreases steadily from 4 nodes on (i.e., 48 cores on), the main cause being an increasing load imbalance,
caused by high idle times of worker tasks/threads (to be addressed in future work). 
However, realistic scientific simulations deal with both, more calculations per particle 
(i.e. detailed physics) and larger numbers of particles, where strong scaling is expected up to thousands of cores. 
We also stress here that the results with $10^6$ particles should not be generalized. 
This relatively low count of particles/core was chosen knowing that the codes will rapidly encounter 
a limit in their strong scaling due to reduced particles count per node with increasing node count, 
and reckoning the limits (in terms of maximum amount of particles that can be handled) of some of the production codes. 
This easily exposed scalability limits of the codes and targets of interest for improvement. Additionally, 
setup and execute experiments with larger amount of particles is much easier when using the mini-app 
than when dealing with its parent codes: express and quick testing is one of the ``raison d'être" of the mini-app.

The experiments show that the selected parent implementations already suffer from load imbalance. 
Acknowledging that imbalance will become an even more important problem on exascale (-ready) machines, the SPH-EXA 
mini-app is being designed to take it into account, already in its early development phases, and addressing it via state-of-the-art, 
scalable dynamic load balancing mechanisms within a single compute node and across massive numbers of nodes. 
\begin{figure}[!htb]
	\centering
		\includegraphics[scale=0.7]{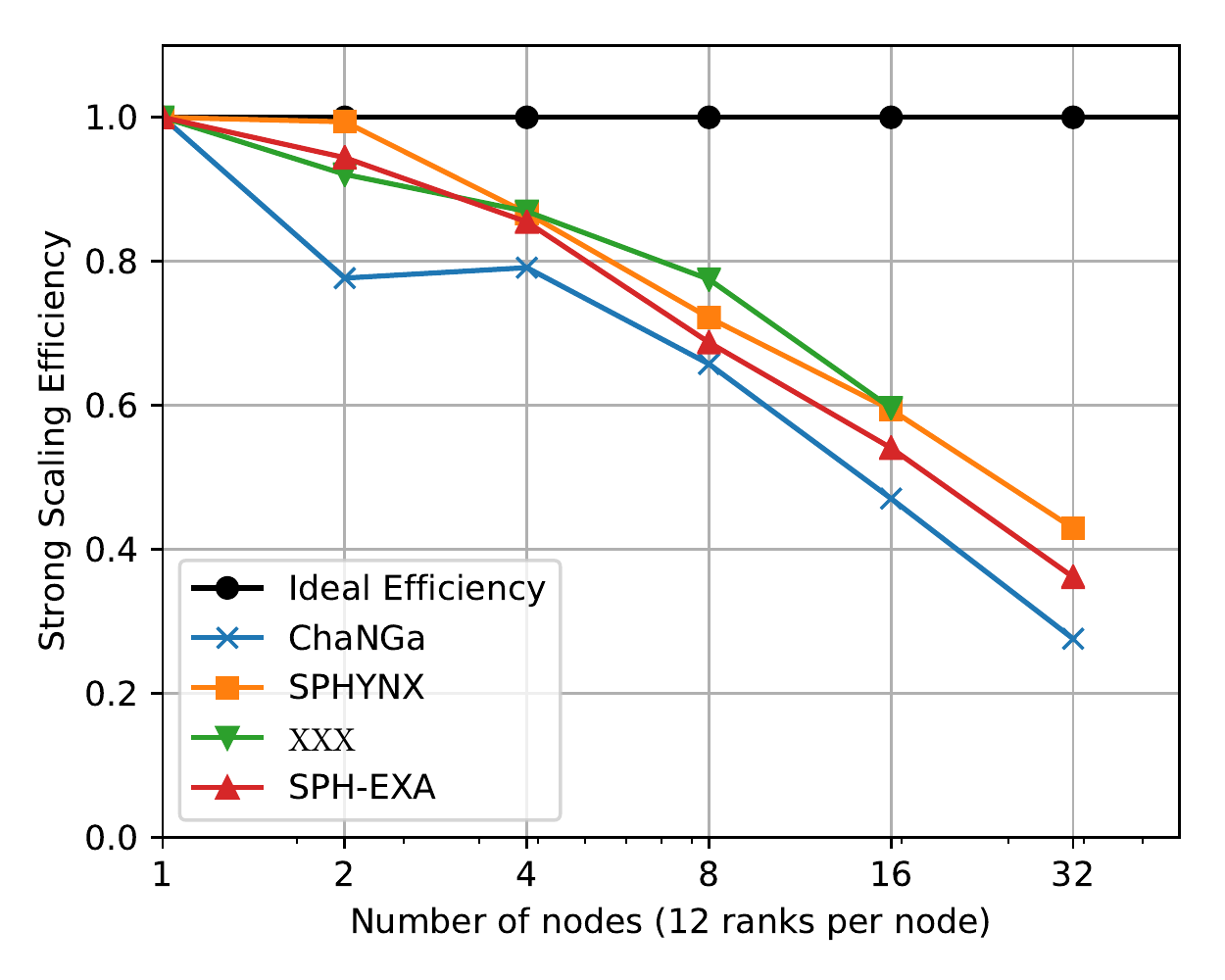}
		\caption{Parallel efficiency of ChaNGa(12 OpenMP tasks and 1 MPI rank per node), SPHYNX (12 OpenMP tasks and 1 MPI rank per node), XXX (12 MPI ranks per node) and SPH-EXA mini-app (12 MPI ranks per node) under strong scaling, in function of the number of nodes used (12 cores per node) on Piz Daint.}
	\label{fig.strong-scaling}
\end{figure}

\subsubsection{Weak scaling evaluation}
\begin{figure}[!htb]
	\centering
		\includegraphics[scale=0.7]{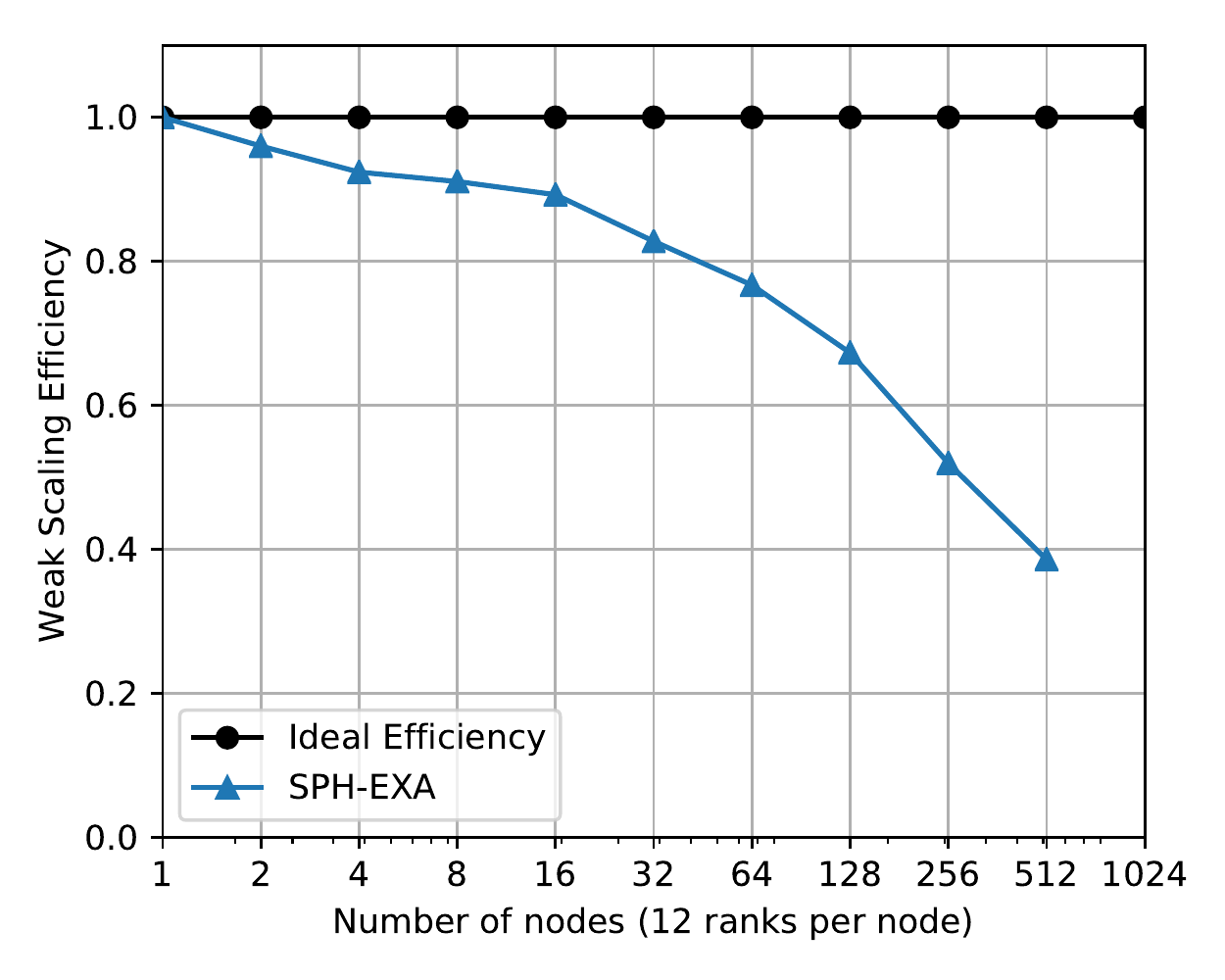}
	\caption{Parallel efficiency of the SPH-EXA mini-app (12 MPI ranks per node) under weak scaling, in function of the number of nodes used (12 cores per node) on Piz Daint.}
	\label{fig.weak-scaling}
\end{figure}

Figure~\ref{fig.weak-scaling} shows the efficiency of the SPH-EXA mini-app  while weak scaling. While performing weak scaling, the problem size assigned to each processing element is constant and additional elements are used to solve a larger problem. The efficiency of weak scaling can be measured as:
\begin{equation}
E = t_s / t_p,\notag
\end{equation}
where $t_s$ is the time of the reference execution and $t_p$ is the time of the execution using $p$ processors. The linear scaling (best case scenario) is achieved if the runtime stays constant while increasing the workload (proportionally to the number of processing elements). 

Compared to the previous strong-scaling experiments where the efficiency drops to $0.3$ upon reaching 32 nodes, weak scaling results show that the proposed SPH-EXA mini-app can scale up to $512$ nodes ($6144$ cores) before the efficiency reaches $0.3$.

\section{Conclusion}

Based on interdisciplinary co-design, we identified the differences and similarities between SPHYNX, ChaNGa and XXX, both in terms of computer science-related features and physical models implemented therein.
This results in a list of selected features, algorithms, and physics modules that are partially incorporated in the SPH-EXA mini-app.

To expose any limitations of the codes that may represent a major performance degradation today or in the future Exascale era, we have compared the codes through a common test cases that serves both as a test, and as a validation and acceptance proof for the mini-app, in the form of reproducible experiment.
The result of this work is a deeper understanding of the three parent SPH codes, with a direct feedback to their developers, that already benefit from this work in terms of unveiling parallelization problems and improving overall scalability. Furthermore, the prototype mini-app has shown promising results at scale, that will be complemented and improved by the features under development.

The SPH-EXA mini-app is expected to enable highly parallelized, scalable, reproducible, portable and fault-tolerant production SPH codes in different scientific domains.
Addressing the performance and scalability challenges of SPH codes requires a versatile collaboration with and support from supercomputing centers, such as CSCS, such that our results can be taken into account for the design of the next generation HPC infrastructures. 

An extensive analysis of the load imbalance in the \mbox{mini-app} both at process and thread level is planned in the future.
Based on the load imbalance analysis, the \mbox{thread-level} load imbalance can be addressed using dynamic loop scheduling techniques implemented in the extended GNU OpenMP runtime library~\cite{Ciorba:2018} or the extended LLVM OpenMP runtime library~\cite{Kasielke:2019}, whereas DLS4LB~\cite{DLS4LB} can be used to address \mbox{process-level} load imbalance. 
Two-level dynamic load balancing ensures improved performance via balanced load execution at both levels~\cite{two-level_DLB}.

In the long-term the goal is to provide a large number of complementary physics modules, thus enabling the mini-app to serve as a production code, that is more diverse and can be used for a broader set of tests than any single of the current parent production codes. Ongoing efforts are the extension to support the integral approach to derivatives and the multipolar expansion. The latter will allow to implement the gravity features thus enabling the mini-app to execute additional simulation scenarios, such as the Evrard collapse and star collision.

Portability is targeted through the porting to HPX, a task-based asynchronous programming model that offers a solution for homogeneous execution of remote and local operations, enabling the exploration of its efficiency and potential on Exascale ready machines.

Faults, errors and failures have become the norm rather than the exception in large-scale systems~\cite{cappello2009, cappello2014, snir2014}. Providing adequate fault-tolerance mechanisms has become mandatory. While checkpointing, rollback and recovery~\cite{chandy1985, elnozahy2002} is the de-facto general-purpose recovery technique to tolerate failures during the execution, optimal multilevel checkpointing can leverage the different storage mediums available to greatly enhance performance. Additional mechanisms to handle silent errors, or silent data corruptions will also be considered~\cite{bautista2016, sridharan2015}.

Reproducibility is at the core of the scientific method, its cornerstone being the ability to independently reproduce and reuse experimental results to prove and build upon them. While the concept of reproducibility has always been part of the science, in certain computational sciences it has long been neglected. A framework has been proposed to support reproducible research~\cite{guerrera2018} and will be analyzed, evaluated, and adopted as in the design of the SPH-EXA mini-app.

\section*{Acknowledgment}
This work is supported in part by the Swiss Platform for Advanced Scientific Computing (PASC) project SPH-EXA~\cite{SPHEXA} (2017- 2020).

The authors acknowledge the support of the Swiss National Supercomputing Centre (CSCS) via allocation project c16, where the calculations have been performed.

Certain computation and simulation results presented here were obtained with XXX, courtesy of XXX Software.


\end{document}